\documentclass[11pt]{article}
\usepackage{amsmath,amsthm,amssymb}
\begin{document}
 \newcommand{\R}{\mathbb{R}}
 \newcommand{\C}{\mathbb{C}}
 \newcommand{\Q}{\mathbb{Q}}
 \newcommand{\Z}{\mathbb{Z}}
 \newcommand{\N}{\mathbb{N}}
 \newcommand{\g}{\;^\circ }
\newcommand{\ggT}{\mbox{\rm ggT}}
\newcommand{\kgV}{\mbox{\rm kgV}}
 \newtheorem{lemma}{Lemma}
 \newtheorem{rem}{Remark}
\newtheorem{thm}{Theorem}
 \newtheorem{cor}{Corollary}
 \newtheorem{bsp}{Example}
 \newtheorem{au}{Aufgabe}
 \newtheorem{defi}{Definition}
\newcommand{\ad}{\mathrm{Ad}}
\newcommand{\spa}{\mathrm{span}}
\newcommand{\Hi}{\mathcal{H}}
\newcommand{\M}{\mathcal{M}}
\newcommand{\W}{\mathcal{S}}
\newcommand{\lO}{\mathcal{O}}
\newcommand{\A}{\mathcal{A}}
\newcommand{\B}{\mathcal{B}}
\newcommand{\lieg}{\mathcal{G}}
\newcommand{\liegf}{\mathfrak{g}}
\newcommand{\liehf}{\mathfrak{h}}
\newcommand{\liesl}{\mathfrak{sl}}
\newcommand{\liegl}{\mathfrak{gl}}
\newcommand{\lieso}{\mathfrak{so}}
\newcommand{\liesu}{\mathfrak{su}}
\newcommand{\liesp}{\mathfrak{sp}}
\newcommand{\abc}{\renewcommand{\labelenumi}{(\alph{enumi})}}
\newcommand{\slim}{\mathrm{s-}\!\!\lim}
\newcommand{\ov}{\overline}
\newcommand{\SO}{\mathrm{SO}}
\newcommand{\sll}{\mathfrak{sl}}
\newcommand{\diag}{\mathrm{diag}}
\newcommand{\trace}{\mathrm{trace}}
\renewcommand{\baselinestretch}{1.7}

\title{Passive States for Essential Observers} 
\author{Robert Strich\\[1mm]
Department of Mathematics, University of Florida}
\date{\today}

\maketitle

\abstract{The aim of this note is to present a unified approach to the results given in \cite{bb99} and \cite{bs04} which also covers examples of models not presented in these two papers (e.g. $d$-dimensional Minkowski space-time for $d\geq 3$). Assuming that a state is passive for an observer travelling along certain (essential) worldlines, we show that this state is invariant under the isometry group, is a KMS-state for the observer at a temperature uniquely determined by the structure constants of the Lie algebra involved and fulfills (a variant of) the Reeh-Schlieder property. Also the modular objects associated to such a state and the observable algebra of an observer are computed and a version of weak locality is examined.} 
\newpage
\section{Introduction}
The algebraic formulation of quantum field theory as introduced by Haag and Kastler (see \cite{ha92}) provides a model independent mathematically rigorous approach to conceptual questions in quantum physics. Within this framework in recent years much work has been done on one of the key problems  one encounters when dealing with quantum field theory on general space-times, namely how to choose physically relevant, fundamental states for a quantum system.  
In \cite{bb99}, \cite{bfs00} and \cite{bs04} it has been shown that if one imposes certain stability conditions on a quantum state for a quantum system on de-Sitter space-time (dS) or Anti-de-Sitter space-time (AdS), then this has strong consequences for the quantum system.
So, for instance (see \cite{bs04} for details), if such a state is passive (\cite{pw78}) with respect to the dynamics of a uniformly accelerated observer in AdS, then this observer sees the given state as an equilibrium state at a certain fixed temperature, the state is invariant under the isometry group of AdS and, what is more, one can deduce weak locality relations among the measurements that the observer in question can perform in his {\it maximal laboratory} and the measurements that can be performed in an {\it opposite} laboratory region.
Similar results were shown to hold in de-Sitter space-time (\cite{bb99}) under  slightly different assumptions, where of course the precise notions of what is meant by {\it maximal laboratory} and {\it opposite} have to be adapted to the respective geometries.
Also, related work has been done for the case of Minkowski space-time in \cite{k02}, but there the author imposes a different set of assumptions.

In this note we are going to generalize these results to quantum systems on {\it a priori} general space-times. One obstacle in this attempt is the absence of concrete geometric information such as the Lie algebraic structure of the symmetry group, which is heavily used in computations in the before mentioned papers. Section 2 tries to overcome this difficulty by introducing a somewhat {\it ad hoc} but useful replacement for the uniformly accelerated observers used in  \cite{bb99} and \cite{bs04} -- called {\it essential} observers.
As already pointed out before, the stability assumptions on a state imply as one of many consequences a fixed Hawking-Unruh temperature (see \cite{un76},\cite{gh77} for details) that a uniformly accelerated  observer (in dS or AdS) finds that state in. In section 3 we show that this temperature is directly related to certain structure constants of the Lie algebra of the symmetry group in question.
In section 4 we provide the basic setup of algebraic quantum field theory  and show how the results of sections 2 and 3 can be applied to obtain similar results as in \cite{bb99} and \cite{bs04}.

\section{Lie group representations and invariant vectors}
Let $G$ be a finite dimensional, connected real Lie group and let $U$ be a strongly continuous, unitary,  faithful representation of $G$ on some Hilbert space $\Hi$. 

Now let's consider the following situation: There is a one-parameter subgroup \linebreak[4]$\{\lambda(t)\}_{t\in\R}\subset G$ and a vector $\phi\in\Hi$ which is invariant under the action of this subgroup, i.e.
\begin{equation}
U(\lambda(t))\phi=\phi\qquad\text{for all}\quad t\in\R.
\end{equation}
In general, this certainly has no implication for the action of the rest of the group $G$ on $\phi$. 
As an example, consider the standard representation of $\SO(3)$ on $L^2(\R^3)$ given by $gf(x)=f(g^{-1}x)$. For every one parameter subgroup of rotations there is an abundance of states $f$ that are invariant under that particular subgroup but not invariant under any other rotation.

But, as for instance shown in \cite{bb99} and \cite{bs04} using direct calculations in the respective Lie groups, for any strongly continuous unitary representation of $\SO(1,n-1)$ or $\SO(2,n-2)$ on some Hilbert space, any vector that is invariant under a boost subgroup 
\begin{equation}\label{boost}
t \mapsto\left(\begin{array}{c c c c c}
\cosh(t) & \sinh(t) & 0 &\cdots& 0\\
\sinh(t) & \cosh(t) & 0 &\cdots & 0\\
0 & 0 & 1 & \cdots &0\\
\vdots &\vdots & \vdots &\ddots &\vdots\\
0 & 0 & 0 & 0 & 1
\end{array}\right)
\end{equation}
must automatically be invariant under the whole group.
We want to give a generalization of the arguments presented there.

Let $\liegf$ be the Lie algebra corresponding to $G$. To any coordinate system in a neighborhood of the identity on our Lie group $G$ we have a set of generators of translations in the coordinate directions $m_1,m_2,\ldots,m_n\in\liegf$ and  we can (via Stone's theorem) find a set  of skewadjoint generators $M_1,M_2,\ldots,M_n$ of $U(G)$ such that $U(\exp(tm_i))=\exp(tM_i)$ for all real $t$ and all $i$. 

 Let $\lieg$ be the real Lie algebra generated by  the  set $\{M_i\}_{1\leq i\leq n}$. Then $\lieg$ consists   of skewadjoint operators acting on some common dense invariant domain of analytic vectors in $\Hi$. Since $U$ is faithful $\lieg$ is isomorphic to $\liegf$.

Now the following is true.
\begin{lemma}\label{l1}
Given a one parameter subgroup $t\mapsto\lambda(t)$ of $G$ let $M\in\lieg$ be its generator i.e. $U(\lambda(t))=\exp(tM)$ for all real $t$. Let furthermore $\phi\in\Hi$ be such that $U(\lambda(t))\phi=\phi$ for all $t\in\R$. Then
\abc
\begin{enumerate}
\item The set $\lieg_\phi=\{N\in\lieg\,|\,\exp(tN)\phi=\phi\,\forall\,t\in\R\}$ is a Lie subalgebra of $\lieg$ containing $M$.
\item If $\ad(M)(N)=\lambda N$ for some $N\in\lieg$ and $\lambda\neq 0$ then $N\in\lieg_\phi$.
\end{enumerate}
\end{lemma}
\begin{proof}
\abc
\begin{enumerate}
\item Using one of the Trotter product formulas, given $N_1,N_2\in \lieg_\phi$ we have
\begin{equation}
\exp\left(t\left(N_1+N_2\right)\right)\phi\\
=\slim_{n\to\infty} \left(\exp(tN_1/n)\exp(tN_2/n)\right)^n\phi=\phi
\end{equation}
for all real $t$. Thus $\lieg_\phi$ is a linear space. Also another Trotter formula
\begin{eqnarray}
\nonumber&&\!\!\!\!\exp(t[N_1,N_2])\phi\\
&=&\!\!\!\!\slim_{n\to\infty}\left(\exp(-tN_1/n)\exp(-N_2/n)\exp(tN_1/n)\exp(tN_2/n)\right)^{n^2}\phi=\phi
\end{eqnarray}
guarantees that $\lieg_\phi$ is a closed under the bracket operation and hence is indeed a Lie subalgebra.
\item Excluding the obvious case $N=0$ we first observe that $\lambda\in\R$ as $-\ov{\lambda} N=[M,N]^*=[N^*,M^*]=-[M,N]=-\lambda N$.
 Now by the Baker-Campbell-Hausdorff formula we have for all real $s,t$
\begin{equation}
\exp(tM)\exp(sN)\exp(-tM)=\exp(se^{t\ad(M)}(N))=\exp(se^{t\lambda}N).
\end{equation}
Set $s=re^{-t\lambda}$ for some fixed but arbitrary $r\in\R$ to get
\begin{equation}
\exp(tM)\exp(re^{-t\lambda}N)\exp(-tM)=\exp(rN).
\end{equation}
Hence if $\lambda>0$ we get, using the fact that $M\in\lieg_\phi$ and that $\exp(tM)$ is unitary of all real $t$:
\begin{align*}
\left|\left|\exp(rN)\phi-\phi\right|\right|&=\lim_{t\to\infty} \left|\left|\exp(rN)\phi-\phi\right|\right|\\
&=\lim_{t\to\infty} \left|\left| \exp(tM)\exp(re^{-t\lambda}N)\exp(-tM)\phi-\phi\right|\right|\\
&=\lim_{t\to\infty} \left|\left|\exp(re^{-t\lambda}N)\phi-\phi\right|\right|\\
&=0.
\end{align*}
The latter follows because of the strong continuity of the representation.
For $\lambda<0$ the same result follows taking the limit $t\to-\infty$.
 Thus we conclude $\exp(rN)\phi=\phi$ for all $r\in\R$, thereby completing the proof.
\end{enumerate}
\end{proof}
Hence it is useful to observe the following.
\begin{lemma}\label{l2}
The following two statements are equivalent for an element $M\in\lieg$:\abc
\begin{enumerate}
\item $M$ together with the set of eigenvectors of $\ad(M)$ in $\lieg$ for nonzero eigenvalues generate $\lieg$ as a Lie algebra;
\item $\ad(M)$ is diagonalizable over $\R$ as a linear map from $\lieg$ to $\lieg$ and the following equation holds:\footnote{as usual we write $[A,B]$ for the linear span of elements of the form $[a,b]$ with $a\in A, b\in B$} 
\begin{equation}\label{e7}
\R M+[M,\lieg]+\left[[M,\lieg],[M,\lieg]\right]=\lieg.
\end{equation}
\end{enumerate}
\end{lemma}
\begin{proof}
If $\ad(M)$ is diagonalizable then $\lieg_*\doteq [M,\lieg]$ is just the span of the eigenvectors belonging to nonzero eigenvalues of $\ad(M)$. Thus $(b)$ implies $(a)$.

On the other hand, if $(a)$ is fulfilled then $\lieg$ is generated as Lie algebra by a set of  eigenvectors of $\ad(M)$ belonging to nonzero eigenvalues together with $M$, which itself is an eigenvector for $\ad(M)$ with eigenvalue $0$.  But if $N_1$ and $N_2$ are eigenvectors for $\ad(M)$ with real eigenvalues $\lambda_1$ and $\lambda_2$ respectively, one has
\begin{equation}
\ad(M)([N_1,N_2])=[M,[N_1,N_2]]=-[N_2,[M,N_1]]-[N_1,[N_2,M]]=(\lambda_1+\lambda_2)[N_1,N_2].
\end{equation}
Thus the commutator $[N_1,N_2]$ is either zero or an eigenvector for the action of $\ad(M)$ with real eigenvalue $\lambda_1+\lambda_2$  which entails that actually $\lieg$ is already spanned by the eigenvectors of $\ad(M)$ as a vector space. Hence $\ad(M)$ is diagonalizable over $\R$.

To prove equation (\ref{e7}) let $N_1=M,N_2,\ldots,N_k$ be a basis of $\lieg$ consisting of eigenvectors of $\ad(M)$ belonging to real eigenvalues $\lambda_1,\lambda_2,\lambda_3,\ldots,\lambda_k$ where $\lambda_i=0$ for $i=1,\ldots,r-1$ and $\lambda_i\neq 0$ for $i\geq r$. 

Then $\lieg_*=[M,\lieg]$ is just the linear span of the set $\{N_i\}_{i\geq r}$.

As $M$ together with the $\ad(M)$-eigenvectors for nonzero real eigenvalues generate $\lieg$ as a Lie algebra, every element in $\lieg$ is a finite linear combination of basic nested commutators of the form $X=[X_1,[X_2,[\ldots[X_{n-1},X_n]\ldots]]]$, where each $X_i$ is either $M$ or one of the $N_j$ for $j\geq r$.  

We have to  show that each such commutator is in $\R\cdot M+[M,\lieg]+[[M,\lieg],[M,\lieg]]=\R\cdot M+\lieg_*+[\lieg_*,\lieg_*]$.

This will be done by induction on the length $n$ of the commutator. For $n=1$ we have $X=X_1$, which equals either $M$ or one of the $N_i$ and thus is in $\R\cdot M+\lieg_*$.

Now let those commutators lie in $\R\cdot M+\lieg_*+[\lieg_*,\lieg_*]$ for all $n\leq n_0$ and consider an $X= [X_1,[X_2,[\ldots[X_{n_0},X_{n_0+1}]\ldots]]]$ of length $n_0+1$.
If $X_1=M$, then $X\in\lieg_*$ and we are done. Otherwise $X_1=N_i$ for some $i\geq r$.
By inductive hypothesis $[X_2,[\ldots[X_{n_0},X_{n_0+1}]\ldots]]$ is a linear combination of elements of the form $\alpha M+\beta N_j+\gamma[N_k,N_l]$ with $j,k,l\geq r$ and real $\alpha,\beta,\gamma$. Thus $X$ is a linear combination of elements of the form 
$$
\alpha[N_i,M]+\beta[N_i,N_j]+\gamma[N_i[N_k,N_l]].
$$
The first summand is in $\lieg_*$, the second lies in $[\lieg_*,\lieg_*]$. Hence we only need to show that $[N_i[N_k,N_l]]\in \R\cdot M+\lieg_*+[\lieg_*,\lieg_*]$

So consider $N=[N_i,[N_k,N_l]]$ with $i,k,l\geq r$. According to the computation above, $N$ is either zero or an eigenvalue for $\ad(M)$ with eigenvalue $\lambda=\lambda_i+\lambda_k+\lambda_l$. If $\lambda\neq 0$ then $N\in\lieg_*$ by definition of $\lieg_*$. Otherwise $\lambda=0$ but then as $\lambda_i\neq 0$ we have $\lambda_j+\lambda_l\neq 0$ and thus $[N_k,N_l]\in\lieg_*$ implying $N\in [\lieg_*,\lieg_*]$. This proves the statement.
\end{proof}

\begin{defi}
If $M\in\lieg$ fulfills any of the equivalent conditions in the lemma, we call it an {\rm essential} element in $\lieg$.
Analogously we call $m\in\liegf$  {\rm essential}, if it fulfills any of the above conditions with $\lieg$ replaced by $\liegf$ and $M$ replaced by $m$.  
\end{defi}
Due to the isomorphism between $\lieg$ and $\liegf$ it follows then in particular, that if $m\in\liegf$ is essential and $U(\exp(tm))=\exp(tM)$, then $M\in\lieg$ is essential and vice versa.
\begin{cor}\label{c1}
If $M\in\lieg_\phi$ is essential, then $\lieg_\phi=\lieg$ and hence $U(\lambda)\phi=\phi$ for all $\lambda\in G$.
\end{cor}
\begin{proof}
According to lemma \ref{l1} and lemma \ref{l2}, $M$ together with  all the eigenvectors of $\ad(M)$ for nonzero real eigenvalues belong to the Lie subalgebra $\lieg_\phi$ of $\lieg$, but also generate $\lieg$ as a Lie algebra. Hence we must have $\lieg_\phi=\lieg$. 

Now (\cite{w03}, \cite{ms03}) for every element $\lambda\in G$ we find $n_1, n_2\in\liegf$ with $\lambda=\exp(n_1)\exp(n_2)$. Hence we find $N_1,N_2\in\lieg$ with $U(\lambda)=\exp(N_1)\exp(N_2)$. Since $N_1,N_2\in\lieg=\lieg_\phi$, we conclude $U(\lambda)\phi=\phi$.
\end{proof}
The following lemma shows that if there is one essential element in a Lie algebra for a Lie group, then there are indeed many of them.
\begin{lemma}
If $M\in\lieg$ is essential then so is $\exp(N)M\exp(-N)\in\lieg$ for all $N\in\lieg$.
\end{lemma}
\begin{proof}
Since $\exp(N)M\exp(-N)$ is the (skew-adjoint) generator of 
\begin{equation}
t\mapsto \exp(N)\exp(tM)\exp(-N),
\end{equation}
 it is indeed in $\lieg$. Also if $K\neq 0$ and $\lambda\neq 0$ such that $[M,K]=\lambda K$, then 
\begin{equation}
[\exp(N)M\exp(-N),\exp(N)K\exp(-N)]=\lambda \exp(N)K\exp(-N)
\end{equation}
and $\exp(N)K\exp(-N)\neq 0$. As $M$ is essential, we conclude that $\exp(N)M\exp(-N)$ together with the eigenvectors of its adjoint action for nonzero eigenvalues generate all of $\exp(N)\lieg\exp(-N)=\lieg$.
\end{proof}
Finally we want to remark, that due to the isomorphism between $\lieg$ and $\liegf$ the following corollary is direct consequence of corollary \ref{c1}.
\begin{cor}
If $m\in\liegf$ is essential and $U(\exp(tm))\phi=\phi$ for all real $t$, then $U(\lambda)\phi=\phi$ for all $\lambda\in G$.
\end{cor}
\subsection{Compact real Lie algebras and essential elements}
We remind the reader that a real semisimple Lie algebra $\liegf$ is called \textit{compact} if its Killing form is negative definite. According to a theorem of Weyl (see for instance \cite[Theorem 2.4]{it}) this is equivalent to the fact that every connected Lie group $G$ having $\liegf$ as Lie algebra is compact. Now the following is true.
\begin{lemma}
Let $\liegf$ be a semisimple, compact real Lie algebra. Then $\liegf$ has no essential elements.
\end{lemma}
\begin{proof}
Assume $m\in\liegf$ is essential. Then $\ad(m)$ is $\R$-diagonalizable. Hence with respect to a suitable basis we have $\ad(m)=\diag(\lambda_1,\lambda_2,\ldots,\lambda_n)$ with $\lambda_i\in\R$. Let $K$ be the Killing form for $\liegf$.

Then $K(m,m)=\trace(\ad(m)^2)=\sum_{i=1}^n\lambda_i^2\geq0$. This contradicts the fact that $K$ is negative definite. Hence there is no such $m$. 
\end{proof}
From this we get the following easy corollary.
\begin{cor}
Neither $\lieso(n)$ nor $\liesu(n)$ have essential elements if $n\geq 2$.
\end{cor}
\subsection{Noncompact real Lie algebras and essential generators}\label{examples}
According to the previous section we will only find examples of real Lie algebras with essential elements among the noncompact ones (or among the non semisimple ones).

In the following we will give  examples of noncompact real Lie algebras $\liegf$ having essential generators.

\begin{description}
\item[$\mathbf{\liegf=\liegl(n,\R)}$] The  Lie algebra $\liegf$ has dimension $n^2$ and generators $e_{\mu,\nu}=E(\mu,\nu)$. Here $E(i,j)$ is the $n\times n$ matrix having a $1$ in row $i$ and column $j$ and zeros elsewhere.

Then each $e_{\nu\nu}$ ($1\leq\nu\leq n$) is essential.
 To see this we observe that 
\begin{equation}\label{cc}
[E(i,j),E(k,l)]=\delta_{jk}E(i,l)-\delta_{li}E(k,j).
\end{equation}
A basis of $\liegf$ consisting of eigenvectors for $\ad(e_{\nu\nu})$ for real eigenvalues is then just the set of all these generators:
$$
\{e_{\mu\rho}\}_{1\leq\mu,\rho\leq n}
$$
Thus $\ad(e_{\nu\nu})$ is $\R$-diagonalizable.
Also $\liegf_*=[e_{\nu\nu},\liegf]=\spa(\{e_{\nu\mu},e_{\mu\nu}\}_{\mu\neq \nu})$ and so
also $e_{\mu\rho}=[e_{\mu\nu},e_{\nu\rho}]\in[\liegf_*,\liegf_*]$ for $\mu,\rho\neq\nu$. Therefore $\R e_{\nu\nu}+\liegf_*+ [\liegf_*,\liegf_*]=\liegf_*$ and hence $e_{\nu\nu}$  is essential.
\item[$\mathbf{\liegf=\liesl(n,\R)}$] The Lie algebra $\liegf$ has dimension $n^2-1$ and generators $e_\nu=E(\nu,\nu)-E(\nu+1,\nu+1)$ ($1\leq\nu\leq n-1$) and $f_{\mu\nu}=E_{\mu,\nu}$ ($\mu\neq\nu$, $1\leq\mu,\nu\leq n$). 

Then each $e_\nu$ is an essential element. 
From the commutator (\ref{cc}) we get
\begin{equation}
\left\{e_\mu\right\}_{1\leq\mu<n}\cup\left\{f_{\mu\rho}\right\}_{\mu\neq\rho, 1\leq\mu,\rho\leq n}
\end{equation}
is a generating set for $\liegf$ consisting of eigenvalues for $\ad(e_\nu)$ (for real eigenvalues). Hence $\ad(e_\nu)$ is $\R$-diagonalizable.

Also $\liegf_*=[e_\nu,\liegf]=\spa\left(\{f_{\nu\mu},f_{\mu\nu}\}_{\mu\neq\nu}\cup \{f_{(\nu+1)\mu},f_{\mu(\nu+1)}\}_{\mu\neq\nu+1}\right)$ and so for $\mu\neq\rho$ (and both $\neq\nu$) $f_{\mu\rho}=[f_{\mu\nu},f_{\nu\rho}]\in[\lieg_*,\lieg_*]$.

But also for $\mu<\nu$ one has $e_\mu+e_{\mu+1}+\ldots+e_{\nu-1}=[f_{\mu\nu},f_{\nu\mu}]\in[\liegf_*,\liegf_*]$ and for  
$\mu>\nu$ one has $e_\nu+e_{\nu+1}+\ldots+e_{\mu-1}=[f_{\nu\mu},f_{\mu\nu}]\in[\liegf_*,\liegf_*]$. 
Thus also all $e_\mu$ belong to $[\liegf_*,\liegf_*]$ and therefore we finally conclude that $\R e_\nu+\liegf_*+[\liegf_*,\liegf_*]=\liegf$. So indeed $e_\nu$ is an essential element.
\item[$\mathbf{\liegf=\liesp(2n,\R)}$] This is a $2n^2+n$ dimensional real Lie algebra with generators $f_{\mu\nu}=E(\mu,\nu+n)+E(\nu,\mu+n)$, $g_{\mu\nu}=E(\mu+n,\nu)+E(\nu+n,\mu)$ and $h_{\mu\nu}=E(\mu\nu)-E(\mu+n,\nu+n)$ where $1\leq\mu,\nu\leq n$. Using the relation (\ref{cc}) one verifies that for instance any $h_{\nu\nu}$ is essential. A (linearly) generating set for  $\liegf$ consisting of eigenvectors for $\ad(h_{\nu\nu})$ for real eigenvalues is just the set of all the above mentioned generators, thus $\ad(h_\nu\nu)$ is $\R$-diagonalizable. Also the space spanned by the eigenvectors for nonzero eigenvalues is
\begin{equation}
\liegf_*=[h_{\nu\nu},\liegf]=\spa(\{g_{\nu\mu}\}\cup\{f_{\nu\mu}\}\cup\{h_{\nu\mu}\}_{\mu\neq \nu}\cup\{h_{\mu\nu}\}_{\mu\neq \nu}).
\end{equation}
Furthermore it then follows that $f_{\mu\rho}=[h_{\mu\nu},f_{\nu\rho}]\in[\liegf_*,\liegf_*]$, $g_{\mu\rho}=[g_{\nu\mu},h_{\nu\rho}]\in[\liegf_*,\liegf_*]$ and $h_{\mu\rho}=[g_{\nu\mu},f_{\nu\rho}]\in[\liegf_*,\liegf_*]$ ($\mu,\rho\neq\nu$). Therefore
 $\R h_{\nu\nu}+\liegf_*+[\liegf_*,\liegf_*]=\liegf$ which means $h_{\nu\nu}$ is essential.
\item[$\mathbf{\liegf=\lieso(1,n).}$] This is the Lie algebra of the identity component of the Lorentz group and $\liegf$ has a generating set  $m_{\mu\nu}$ with $0\leq \mu,\nu\leq n$ fulfilling the Lie algebra relations
\begin{equation}\label{cr}
[m_{\mu\nu},m_{\rho\sigma}]=g_{\mu,\rho}m_{\nu\sigma}+g_{\nu\sigma}m_{\mu\rho}-g_{\mu\sigma}m_{\nu\rho}-g_{\nu\rho}m_{\mu\sigma}
\end{equation}
where $g=\diag(1,-1,-1,\ldots,-1,-1)$ and $m_{\mu\nu}=-m_{\nu\mu}$. Then any of the elements $m_{0\nu}$ with $1\leq \nu\leq n$ is essential. A generating set for $\liegf$ of eigenvectors for real eigenvectors of $\ad(m_{0\nu})$ is given by
\begin{equation}
\{m_{0\mu}\pm m_{\nu\mu}\}_{\mu\neq 0,\nu}\cup\{m_{\mu\rho}\}_{\mu,\rho\notin\{0,\nu\}}\cup\{m_{0\nu}\}.
\end{equation} 
Thus $\ad(m_{0\nu})$ is $\R$-diagonalizable.

Also $\liegf_*=[m_{0\nu},\liegf]=\spa\left(\{m_{0\mu}\pm m_{\nu\mu}\}_{\mu\neq 0,\nu}\right)=\spa\left(\{m_{0\mu}, m_{\nu\mu}\}_{\mu\neq 0,\nu}\right)$ and as furthermore $[m_{0\mu},m_{0\rho}]=m_{\mu\rho}$ for $\mu,\rho\notin\{0,\nu\}$ we have $\R m_{0\nu}+\liegf_*+[\liegf_*,\liegf_*]=\liegf$.
\item[Poincar\'e algebra.] Here $\liegf$  is the Lie algebra of the identity component of the Poincar\'e group $G=\SO(1,n)^+\ltimes\R^{n+1}$ and it has in addition to the generators $m_{\mu\nu}$ above the   translation generators $p_\mu$ for $0\leq\mu\leq n$ with the additional Lie algebra relations
 \begin{align}
[p_\mu,p_\nu]&=0\\
[m_{\mu\nu},p_\sigma]&=g_{\mu\sigma}p_\nu-g_{\nu\sigma} p_\mu.
\end{align}
Still the elements $m_{0\nu}$ with $1\leq \nu\leq n$ are essential. We can simply prolong the list of eigenvectors for $\ad(m_{0\nu})$ generating $\liegf$ from above by $\{p_0\pm p_\nu\}\cup\{p_\mu\}_{\mu\notin\{0,\nu\}}$ - hence $\ad(m_{0\nu})$ is again diagonalizable and 
\begin{equation}
\liegf_*=\spa\left(\{m_{0\mu}, m_{\nu\mu}\}_{\mu\neq 0,\nu}\cup\{p_0,p_\nu\}\right).
\end{equation}
As also $[m_{0\mu},p_0]=p_\mu$ we again get $\R m_{0\nu}+\liegf_*+[\liegf_*,\liegf_*]=\liegf$.
\item[$\mathbf{\liegf=\lieso(p,q).}$] The Lorentz algebra example above can be easily generalized to any Lie algebra $\lieso(p,q)$ with $p,q\geq 1$.  If the generators are labelled as before by $m_{\mu\nu}$ with $1\leq \mu,\nu\leq p+q$ but now $g=\diag(1,1,\ldots,1,-1,\ldots,-1)$ with $p$ entries $1$ and $q$ entries $-1$ then every element $m_{\mu\nu}$ with $\mu\leq p$ and $\nu\geq p+1$ will be essential by an analoguous calculation as above.
\end{description}
\section{$\beta$-KMS-states}

In \cite{bb99}, \cite{bfs00} and  \cite{bs04} the authors show how the geometry of the de Sitter space-time and that of Anti-de-Sitter space-time, in particular the specific commutation relations in the corresponding  symmetry groups, determine the value $\beta$ for a $\beta$-KMS-state (see \cite{ha92}) with respect to the dynamics given by a boost subgroup uniquely in each of the two space-times. Their results rely heavily on concrete calculations in the corresponding Lie algebras. By generalizing their arguments, we show in the following that the value of $\beta$ is directly related to certain structure constants in the Lie algebra of the isometry group of the given general space-time. 
\begin{thm}\label{t}
Let $\phi\in\Hi$ be a $\beta$-KMS-state for the dynamics given by a skew-adjoint generator $M$ on a von Neumann algebra $\A\subset B(\Hi)$ and let $\beta>0$. Let furthermore $N$ be skew adjoint such that
\abc
\begin{enumerate}
\item $\ad(M)(N)=\lambda N$ for some nonzero $\lambda$ and $N$;\item there is a sub-algebra $\B\subset\A$ such that 
\begin{equation}
\exp(tN)\exp(rM)\B\exp(-rM)\exp(-tN)\subset \A
\end{equation}
 for  $|r|+|t|<\delta$ for some $\delta>0$, and $\phi$ is cyclic for $\B$.
\end{enumerate} 
Then $\beta=\frac{2\pi}{|\lambda|}$. 
\end{thm}
\begin{proof}
As $\phi$ is cyclic for $\B\subset \A$, $\phi$ is also cyclic for $\A$. To see that it is also separating consider $A\in\A$ such that $A\phi=0$. Now as $\phi$ is a KMS state there is a function $f$ continuous in the complex strip $S_\beta\doteq \{z\,|\,0\leq \Im(z)\leq \beta\}$ and analytic in the interior of that strip such that for real $t$ and $B,C\in\A$
\begin{equation}
f(t)=(\phi,C^*A\exp(tM)B\phi)\,\,\,\mathrm{and}\,\,\,f(t+i\beta)=(\phi,B\exp(-tM)C^*A\phi)=0.
\end{equation}
Hence $f$ vanishes everywhere in $S_\beta$, and we have in particular $f(0)=(C\phi,AB\phi)=0$. As this holds for arbitrary $B,C\in\A$ and $\phi$ is cyclic for $\A$, we get $A=0$.

As $\phi$ is cyclic and separating for $\A$, we can consider the modular operator $\Delta$ and the modular conjugation $J$ associated with the pair $(\A,\phi)$. 

The fact that the adjoint action of $\exp(tM)$ leaves $\A$ invariant and fulfills the KMS property entails (\cite{pw78}) that $\exp(tM)\phi=\phi$ for all $t$ and
\begin{equation}
\Delta^{it}=\exp(-\beta t M).
\end{equation}
Consequently, we can also compute that for all $A\in \A$
$$
JA\phi=J\left(J\Delta^{\frac{1}{2}}\right)A^*\phi=\exp\left(\frac{i\beta}{2}M\right)A^*\phi.
$$
As a consequence of the commutation relation $(a)$, we have for all real $t,s$:
\begin{equation}
\exp(sM)\exp(tN)=\exp(t\exp(\lambda s)N)\exp(sM),
\end{equation}
and we also know (Lemma \ref{l1}) that $\lambda\in\R$ and $\exp(tN)\phi=\phi$ for all $t\in\R$.

Now pick any $B\in \B$. Then one  has for any $\psi\in\Hi$:
\begin{equation}\label{e17}
(\psi,\exp(sM)\exp(tN)B\exp(-tN)\phi)=(\psi,\exp(t\exp(\lambda s)N)\exp(sM)B\phi).
\end{equation}
By assumption, $\exp(tN)B\exp(-tN)\in\A$ for $|t|<\delta$ and hence we conclude that \linebreak[4] $\exp(tN)B\exp(-tN)\phi$ is in the domain of $\Delta^{1/2}=\exp(\frac{i\beta}{2} M)$. Thus the left hand side can be analytically continued in $s$ into the strip $S_{\frac{\beta}{2}}$ and has a continuous limit at the upper end of that strip.

For the right hand side observe that there is a dense set of Nelson vectors $\psi$ for which $z\mapsto \exp(zN)\psi$ can be analytically continued in $z$ inside a ball $B(0,\rho)\subset \C$. Thus for a Nelson vector $\psi$ the function $s\mapsto \exp(-t\exp(\lambda s)N)\psi$ can be analytically continued  in $s$ into the ball $B(0,\rho(t))$ with  $\rho(t)=\log(\rho\cdot|t|^{-1})\cdot\lambda^{-1}\to\infty$ as $|t|\to 0$.

As also $s\mapsto \exp(sM)B\phi$ allows an analytic continuation into $S_{\frac{\beta}{2}}$ by the same argument as above, we deduce that both sides of the last equation can be analytically continued into the region $S_{\frac{\beta}{2}}\cap B(0,\rho(t))$ with  continuous boundary values and are hence equal there. Thus for sufficiently small $|t|$ we can set in particular $s=\frac{i\beta}{2}$ to get the equality
\begin{equation}
\Big(\psi,\exp\Big(\frac{i\beta}{2}M\Big)\exp(tN)B\exp(-tN)\phi\Big)\!\!=\!\!\Big(\psi,\exp\Big(t\exp\Big(\frac{i\beta\lambda}{2} \Big)N\Big)\exp\Big(\frac{i\beta}{2}M\Big)B\phi\Big).
\end{equation}
 This is equivalent to 
\begin{equation}
(\psi,J\exp(tN)B^*\exp(-tN)\phi)=\Big(\psi,\exp\Big(t\exp\Big(\frac{i\beta\lambda}{2} \Big)N\Big)JB^*\phi\Big).
\end{equation}
Now since this is true for a dense set of vectors $\psi$ and since $\phi$ is cyclic for $\B$ by assumption, we get
\begin{equation}\label{eq22}
J\exp(tN)=\exp\Big(t\exp\Big(\frac{i\beta\lambda}{2} \Big)N\Big)J
\end{equation}
for small $|t|$. After iterating this equation suitably often, we see that it actually holds for all real $t$.

As $J$ is anti-unitary and $\exp(tN)$ is unitary it then follows that $\exp(\frac{i\beta\lambda}{2})\in\R$, i.e. $\beta=\frac{2\pi k}{|\lambda|}$ for some positive integer $k$.

Now suppose $k\geq 2$. Setting $B=\exp(rM)C\exp(-rM)$ for $C\in\B$ in equation (\ref{e17}), we see that for $|r|+|t|<\delta$  both sides of the equation 
  allow an analytic continuation in $s$ into the region $S_{\frac{\beta}{2}}\cap B(0,\rho(t))$.
 Setting first $s=\frac{\pi i}{|\lambda|}<\frac{\beta}{2}$ yields 
\begin{equation}
\Big(\!\psi,\exp\Big(\frac{\pi i}{|\lambda|}M\Big)\!\exp(tN)\!\exp(rM)B\exp\phi\Big)\!\!=\!\!\Big(\!\psi,\exp(-tN)\!\exp\Big(\frac{\pi i}{|\lambda|}M\Big)\!\exp(rM)B\phi\Big)
\end{equation}
for small enough $|t|$ and $|r|$. Again, as this holds for a dense set of vectors $\psi$, we have
\begin{equation}
\exp\Big(\frac{\pi i}{|\lambda|}M\Big)\exp(tN)\exp(rM)B\phi=\exp(-tN)\exp\Big(\frac{\pi i}{|\lambda|}M\Big)\exp(rM)B\phi
\end{equation}
for small $|t|$ and $|r|$. Now consider any compact Borel set $\Delta$ and let $P(\Delta)$ be the projection onto the corresponding spectral subspace of the (selfadjoint) operator $iM$. Then multiplying the previous equation with $P(\Delta)$ from the left gives
\begin{equation}
\exp\Big(\frac{\pi i}{|\lambda|}M\Big)P(\Delta)\exp(tN)\exp(rM)B\phi\!=\!P(\Delta) \exp(-tN)\exp\Big(\frac{\pi i}{|\lambda|}M\Big)\exp(rM)B\phi.
\end{equation}
Since $\exp\Big(\frac{\pi i}{|\lambda|}M\Big)P(\Delta)$ is a bounded operator and since (as $\frac{\beta}{2}\geq \frac{2\pi}{|\lambda|}$)  $B\phi$ is in the domain of  $\exp\Big(\frac{2\pi i}{|\lambda|}M\Big)$, we can again continue both vector-valued sides analytically in $r$ into $S_{\frac{\pi i}{|\lambda|}}$. Hence (by the Edge-of-the-Wedge theorem) the last equality does not only hold for small $|r|$, but for all real $r$ and small $|t|$. This implies
\begin{equation}
\exp\Big(\frac{\pi i}{|\lambda|}M\Big)P(\Delta)\exp(tN)P(\Delta)B\phi=P(\Delta) \exp(-tN)\exp\Big(\frac{\pi i}{|\lambda|}M\Big)P(\Delta)B\phi.
\end{equation}
Now, as we are dealing only with bounded operators, the fact that $\phi$ is cyclic for $\B$ entails
\begin{equation}
\exp\Big(\frac{\pi i}{|\lambda|}M\Big)P(\Delta)\exp(tN)P(\Delta)=P(\Delta) \exp(-tN)P(\Delta)\exp\Big(\frac{\pi i}{|\lambda|}M\Big)
\end{equation}
for small $|t|$. Hence $P(\Delta)\exp\Big(\frac{2\pi i}{|\lambda|}M\Big)P(\Delta)$ commutes with $P(\Delta)\exp(tN)P(\Delta)$ for small $|t|$. This entails (\cite[Lemma 5.6.13, 5.6.17]{kr83}) that the spectral projections of the selfadjoint operators $P(\Delta)iMP(\Delta)$ and $P(\Delta)iNP(\Delta)$ commute. As $\Delta$ was arbitrary this in particular implies that $M$ and $N$ commute, contradicting the assumptions.

Consequently we must have $\beta=\frac{2\pi}{|\lambda|}$.
\end{proof}

\section{Application to AQFT}
\subsection{Basic Setup}
We will now show how the previous results can be applied in quantum field theoretic problems. We will make use of the algebraic formulation of quantum field theory as introduced by Haag and Kastler (see \cite{ha92} for more details).

In particular we will be considering an $n$-dimensional manifold $\M$ together with a Lorentzian metric that models our space-time. 
Whereas a generic space-time $\M$ will have a trivial isometry group, for our approach it is crucial that  $\M$ has indeed nontrivial symmetries. We consider a connected subgroup $G$ of the isometry group of $\M$ and assume that it is strongly continuously, unitarily and faithfully represented on some separable Hilbert space $\Hi$ via the representation $U$.

The observables of the theory form an isotonous net of von Neumann algebras $\A(\lO)$ indexed by open subsets $\lO\subset \M$, i.e. we have an assignment $\lO\mapsto\A(\lO)$ such that $\lO_1\subset\lO_2$ implies $\A(\lO_1)\subset\A(\lO_2)$. The (global) observable algebra $\bigvee_{\lO\subset\M} \A(\lO)\doteq \left(\bigcup_{\lO\subset\M} \A(\lO)\right)''$ is denoted by $\A$.  Also $G$ is assumed to act covariantly upon the net, i.e. for every $g\in G$ and every open $\lO\in\M$ we have
\begin{equation}
U(g)\A(\lO)U(g)^*=\A(g\lO).
\end{equation}

\subsection{Observers and Wedges}
We are looking at observers travelling along worldlines generated by a one-paramter group of isometries. To be precise let $\{\lambda(t)\}_{t\in\R}$ be a one-parameter subgroup of $G$. If for some $x\in\M$ the curve $t\mapsto \lambda(t)x$ is timelike everywhere, we regard it as a possible worldline of an observer.

Let $W(\lambda)$ be the open set of all $x$ for which $t\mapsto \lambda(t)x$ is a timelike curve. The connected component of $W(\lambda)$ that contains the given worldline, i.e. the set of all neighboring worldlines, will be called the {\it wedge} $W(\lambda,x)$ associated to the observer, respectively associated to the worldline. This is typically the set of events that can influence or can be influenced by our observer. In any case we regard $W(\lambda,x)$ as the maximal localization region of observables that can be measured by the observer. 

In Minkowski space-time, for instance, wedges for the boost-subgroup (\ref{boost}) are precisely the wedge shaped regions  $W_R=\{x\in\R^4\,|\,|x_0|<x_1\}$ and $W_L=-W_R$. 

A technical requirement on the size of the wedges and the size of $G$ is the following: Let $\W$ be the set of $\lO\subset \M$ for which $\bigvee_{g\in G} \A(g\lO)=\A$. Also we call an inclusion $\lO_1\subset\lO_2$ of open subsets of $\M$ {\it proper} if there is an open neighborhood $N$ of $1$ in $G$ such that $N\lO_1\subset\lO_2$. Then we require
\begin{quote}
{\bf(WA)} (Weak Additivity): Each wedge $W(\lambda,x)$ has a proper subset in $\W$.
\end{quote}
In Minkowski space-time this holds under very general assumptions \cite{sw87}. It even holds in models in which the local algebras localized in sufficiently small regions are trivial, e.g. \cite{bms00}.
\begin{lemma}
The set of wedges is invariant under the action of $G$. If $W(\lambda,x)$ fulfills $(WA)$ then so does $gW(\lambda,x)$ for all $g\in G$.
\end{lemma}
\begin{proof}
When $t\mapsto \lambda(t)x$ is a timelike curve then so is $t\mapsto g\lambda(t)x$ for all $g\in G$. This implies the result.
\end{proof}

\subsection{States}
One of the key problems in  quantum field theory on general space-times is to pick states of interest out of the abundance of possible states.
In the following we will introduce a list of properties which can be used to characterize fundamental states for quantum systems (\textit{vacua}). A similar approach was taken in \cite{bfs00}. We do not assume our state to have all these properties; instead we will show in the following sections how these properties are related in our special situation.

First of all, we can assume that such  a state  will be represented by a normalized vector $\Omega\in\Hi$ (by considering the GNS-representation associated to our state). We can also assume that $\Omega$ is cyclic for $\A$ since otherwise we could just restrict ourselves to a smaller Hilbert space.

 Furthermore it is well known, that the vacuum state in quantum field theories constructed on Minkowski, de-Sitter and Anti-de-Sitter space-time is invariant under  symmetries of the respective space-times. Therefore it is in general desirable for such a fundamental vector state to be invariant under isometries. Hence we introduce the following notion:
\begin{quote}
{\bf(I)} (Invariance): $\qquad U(g)\Omega=\Omega$ for all $g\in G$.
\end{quote}
Also an observer freely falling along a worldline described above should see this potential vacuum $\Omega$  as energetically stable in the sense that the expected value of the energy in this state is minimal among the energy expectations in small perturbations of $\Omega$. The mathematical description of this property of a state is as follows (see also \cite{pw78}):
\begin{quote}
{\bf{(P)}} (Passivity): $\Omega$ is a passive state for observers travelling along certain worldlines $t\mapsto \lambda(t)x$. This means that for all unitary $V\in\A(W(\lambda,x))$ and for the selfadjoint generator $M$ of $U(\lambda(t))$ we have
\begin{equation}
(V\Omega,MV\Omega)\geq (\Omega,M\Omega).
\end{equation}
\end{quote}
We will always make clear what exactly we mean by \textit{certain worldlines} when we impose the passivity condition on a state.

Another requirement for $\Omega$ (and for the net of observables) is  that $\Omega$ is fundamental in the sense that each other state can be at least approximatively prepared out of $\Omega$ by operations performed just in some open region $\lO$ properly contained in the maximal laboratory $W(\lambda,x)$ of an observer. Mathematically speaking this is the
\begin{quote}
{\bf{(RS)}} (Reeh-Schlieder-property): Each wedge contains properly an open $\lO$ such that $\Omega$ is cyclic for $\A(\lO)$, i.e.
\begin{equation}
\overline{\A(\lO)\Omega}=\Hi.
\end{equation}
\end{quote}
The Reeh-Schlieder-property was first shown under very general assumptions to be a feature of vacuum states for quantum field theories on Minkowski space-time in \cite{rs61}. In \cite{bb99} and \cite{bs04} the Reeh-Schlieder-property was shown to be a consequence of certain stability conditions on a state in de-Sitter and Anti-de-Sitter space-time.
 
Another property of states describing pure thermodynamical phases (see \cite{pw78}, \cite{hkt74}) is the following. It describes the fact that in a pure phase in mean the correlation between observables respectively localized in two regions decays suitably fast as a function of their timelike separation with respect to the dynamics given by $M$.
\begin{quote}
{\bf{(WM)}} (Weak Mixing): $\Omega$ is weakly mixing for an observer travelling along $t\mapsto\lambda(t)x$ if for all $A,B\in\A(W(\lambda,x))$ the expression 
\begin{equation}
\frac{1}{T}\int_0^T\left((\Omega, \ad(\exp(itM))(A)B\Omega)- (\Omega, \ad(\exp(itM))(A)\Omega)(\Omega,B\Omega)\right)dt
\end{equation}
vanishes in the limit $T\to\infty$.
\end{quote}

Finally, a state $\Omega$ is called \textit{central}  for some observer travelling along $t\mapsto \lambda(t)x$ if $(\Omega,AB\Omega)=(\Omega,BA\Omega)$ for all $A,B\in \A(W(\lambda,x))$. In this case either $\Omega$ is annihilated by most of the elements in $\A(W(\lambda,x))$ or this algebra is of finite type (see \cite{kr88} for details). These are (from the view of quantum field theory) pathological circumstances that we want to avoid. Therefore one  introduces the following notion:
\begin{quote}
{\bf{(NC)}} (Noncentrality): $\Omega$ is not central for certain observers.
\end{quote}
Again, it will be made clear with respect to which observer we want $\Omega$ to be noncentral when we impose this condition on a state. 
\subsection{Invariance and Reeh-Schlieder Property}
We now want to investigate some relations among these properties when we deal with subgroups generated by essential elements. The arguments presented, as well as the idea of taking the assumption of passivity as a starting point of the investigation, were first published for the special case of Anti-de-Sitter space-time in \cite{bfs00} and \cite{bs04}.
\begin{thm}
Let $\Omega$ fulfill properties (P), (WM) and (NC) for an observer travelling along $t\mapsto \lambda(t)x$ with $U(\lambda(t))=\exp(tM)$ for some (skew-adjoint) essential $M$. Then
\abc
\begin{enumerate}
\item $\Omega$ fulfills $(I)$;
\item if $-iM$ is not a positive operator, then $\Omega$ fulfills $(RS)$ for all wedges $gW(\lambda,x)$ with $g\in G$ as well. 
\end{enumerate}
\end{thm} 
\begin{proof}
\abc
\begin{enumerate}
\item Using deep results of Pusz and Woronowicz (\cite{pw78}), the passivity and the weak mixing property of $\Omega$ entail that $M\Omega=0$ and $\Omega$ is either a ground state or a KMS-state at some inverse temperature $\beta\geq 0$ for $M$. Thus Corollary \ref{c1} implies that $\Omega$ is invariant under the whole group action. 

\item It suffices to show the result for $W(\lambda,x)$, since 
\begin{equation}
\A(gW(\lambda,x))=U(g)\A(W(\lambda,x))U(g)^*.
\end{equation}
Since $W(\lambda,x)$ fulfills (WA), there is an open $\lO\in\W$ properly included in $W(\lambda,x)$, i.e. there is an open neighborhood $N$ of $1\in G$ such that $N\lO\subset W(\lambda,x)$. The preimage of $N$ under the continous product map $G\times G\to G$ contains an open rectangle $L_1\times L_2$ containing $(1,1)$. Then $L\doteq L_1\cap L_2$ is a second neighborhood of $1\in G$ such that $L^2\subset N$. Set $K\doteq L\cap N$.

Then $K\lO\subset N\lO\subset W(\lambda,x)$ and $K(K\lO)=K^2\lO\subset N\lO\subset W(\lambda,x)$. Hence $K\lO$ is properly included in $W(\lambda,x)$.

 Consider a vector $\phi\in\Hi$ such that 
\begin{equation}
(\phi,A\Omega)=0
\end{equation}
for all $A\in\A(K\lO)$. We are going to show that $\phi=0$.\\
Pick a $B\in \A(\lO)$ and any $g\in K\cap N^{-1}$. Then for small $|t|<\epsilon$ we have
\begin{equation}
\ad(U(g\lambda(t)g^{-1}))(B)\in \A(K\lO) 
\end{equation} 
Hence we have that
\begin{equation}
f(t)=(\phi, U(g \lambda(t) g^{-1}) B \Omega)=(\phi, U(g)\exp(tM) U(g^{-1}) B \Omega)
\end{equation}
vanishes on $|t|<\epsilon$. Since $-iM$ is not a positive operator, $\Omega$ is not a ground state for the dynamics $\exp(tM)$. Hence according to  the results of Pusz and Woronowicz mentioned above,  $\Omega$ is a KMS state for some inverse teperature $\beta\geq 0$. In fact as the representation $U$ is assumed to be faithful and $\Omega$  is noncentral, we must have $\beta>0$. Furthermore we know that for $g\in K\cap N^{-1}$ we have $U(g^{-1})BU(g)\in \A(g^{-1}\lO)\subset\A(N\lO)\subset\A(W)$.  Thus the function $f:t\mapsto  U(g)\exp(tM) U(g^{-1}) B \Omega$ is a boundary value of an analytic vector valued function in a strip in the complex plane, hence it vanishes everywhere in that strip. In particular, $f$ vanishes for all $t\in \R$.\\

Hence we have
\begin{equation}
( U(g \lambda(t) g^{-1})\phi, B \Omega)=0
\end{equation}
for all $t\in\R$. Repeating the same argument several times we get that 

\begin{equation}
( U(g_1 \lambda(t_1) g_1^{-1})U(g_2 \lambda(t_2) g_2^{-2})\ldots U(g_k \lambda(t_k) g_k^{-1})\phi, B \Omega)=0
\end{equation}
for all $t_i\in\R$ and $g_i\in K\cap N^{-1}$. 

Now  we prove the following small lemma.
\begin{lemma}
Let $N\subset G$ be any open neighborhood of the identity in $G$. 
Then the strong operator closure $H$ of the group generated by the unitaries $U(\lambda\lambda(t)\lambda^{-1})$ for $t\in\R$, $\lambda\in N$ coincides with all of $U(G)$. 
\end{lemma}
\begin{proof}
Let $\lieg'=\{K\in\lieg\,|\, \exp(tK)\in H,\,\forall t\in\R\}$.
Then using Trotter formulae as in the Lemma 1 $(a)$, we readily see that $\lieg'$ is a Lie subalgebra of $\lieg$.

Hence  due to the essentiality of $M$, it suffices to show that $M\in\lieg'$ and $K\in\lieg'$, if $[M,K]=\lambda K$ for real nonzero $\lambda$ and $K\in\lieg$.
While the first is obvious, for the second we argue as follows:
As $N$ is an open neighborhood of $1$ in $G$, we find $n_0\in\N$ such that $\exp(K/n)\in N$ and $\exp(-K/n)\in N$  for all $n\geq n_0$ and hence
$$
\exp(-K/n)\exp(-tM/n)\exp(K/n)\in H
$$
for all $n\geq n_0$ and all real $t$. Thus we have (again by the Trotter formula)
\begin{align*}
\slim_{n\to\infty}\left(\exp(-K/n)\exp(-tM/n)\exp(K/n)\exp(tM/n)\right)^{n^2}&=\exp([tM,K])\\
&=\exp(t\lambda K)\in H
\end{align*}
for all real $t$ and as $\lambda\neq 0$. So we have $K\in\lieg'$, which finishes the proof of the lemma.
\end{proof}
Hence we conclude
\begin{equation}
(\phi, U(g)B\Omega)= ( U(g^{-1})\phi, B \Omega)=0
\end{equation}
for all $g\in G$.
 From this and the fact that $B\in \A(\lO)$ was arbitrary, we finally deduce that then also 
\begin{equation}
\left(\bigvee_{g\in G} U(g) \A(\lO)U(g^{-1})\right)\Omega=\left(\bigvee_{g\in G} \A(g\lO)\right)\Omega=\A\Omega
\end{equation} 
is perpendicular to $\phi$ which implies $\phi=0$ as $\Omega$ is cyclic for $\A$.
\end{enumerate}
\end{proof}
\subsection{Modular Objects and Unruh-Temperature}
In  a classic paper (\cite{bw75a}) Bisognano and Wichmann showed that the modular objects associated to a vacuum state and the algebra of observables of a wedge region in Minkowski space-time (generated by Wightman fields) act geometrically upon the net of observable algebras. This result has been extended to various other space-times (\cite{bb99},\cite{bs04},\cite{ms96}). Also, in light of these results, the property of a state (and a net of observable algebras), that certain modular objects act geometrically, were proposed as a selection criterion for physically relevant states (\cite{bs93},\cite{bdfs98}).
In this section we show that also under our general assumptions the modular objects have a geometric interpretation and that the corresponding Unruh-Temperature can be determined.

The following holds as long as the Lie group $G$ has at least dimension $2$.
\begin{thm}
Let $\Omega$ fulfill properties (P), (WM) and (NC) for an observer travelling along $t\mapsto \lambda(t)x$ with $U(\lambda(t))=\exp(tM)$ for some (skew-adjoint) essential $M$ such that $-iM$ is not positive. Then
\abc
\begin{enumerate}
\item there is a nonzero eigenvalue $\lambda$ for the adjoint action of $M$ on $\lieg$; all such eigenvalues have the same modulus  and $\Omega$ is a $\frac{2\pi}{|\lambda|}$-KMS state for the dynamics $\exp(tM)$ on $\A(W(\lambda,x))$;
\item $\Omega$ is cyclic and separating for $\A(W(\lambda,x))$ and hence for each $\A(gW(\lambda,x))$ for all $g\in G$.
 The modular operators for the pair $(\A(W,\lambda),\Omega)$ are given as
 \begin{equation}\label{44}
 \Delta^{it}=\exp(-\frac{2\pi}{|\lambda|} t M)\qquad\text{and}
\end{equation}
\begin{equation}\label{eq41}
JA\Omega=\exp\left(\frac{i\pi}{\lambda}M\right)A^*\Omega.
\end{equation}
Furthermore the commutation relations of $J$ with the group representations are fixed as follows: $J$ commutes with $\exp(tN)$ if $[M,N]=0$ and if $[M,N]=\pm \lambda N$ then
\begin{equation}\label{eq42}
J\exp(tN)=\exp(-tN)J.
\end{equation}
\end{enumerate}
\end{thm}
\begin{proof}
\abc
\begin{enumerate}
\item As $M$ is essential and $\lieg$ has dimension greater than $2$ there must be some nonzero $N\in \lieg$ and some nonzero $\lambda$ with $[M,N]=\lambda N$. As seen before, properties (P), (WM) and (NC) entail that $\Omega$ is a $\beta$-KMS state for some $\beta>0$. Also, from the previous theorem the Reeh-Schlieder property holds for $\A(W(\lambda,x))$, and hence there is an $\lO$ properly included in $\A(W(\lambda,x))$ such that $\Omega$ is cyclic for $\A(\lO)$. 

Therefore all the assumptions for theorem \ref{t} are fulfilled, and we conclude that indeed $\beta=\frac{2\pi}{|\lambda|}$ and hence $|\lambda|$ is uniquely given.
\item This is proved in the first part of the proof of theorem \ref{t}. The commutation relations follow from equation (\ref{eq22}) by plugging in $\beta=\frac{2\pi}{|\lambda|}$.
\end{enumerate}
\end{proof}
In the special case of Minkowski space-time, the derived condition (\ref{44}) is known as \textit{modular covariance}. With modular covariance as one of the assumptions, the authors of \cite{bgl94} derive a representation of the Poincar\'e group which acts covariantly upon the net.
\subsection{Weak Locality}
The stated assumptions on $\Omega$ seem not to suffice to deduce strong locality relations in general, see for instance the examples constructed in \cite{bs04} and \cite{bs07} on AdS and Minkowsi space-time respectively. But one can at least formulate the following result on weak locality. Similar results in the special case of Anti-de-Sitter space-time had first been published in \cite{bs04}.
\begin{thm}
Let $\Omega$ fulfill properties (P), (WM) and (NC) for an observer travelling along $t\mapsto \lambda(t)x$ with $U(\lambda(t))=\exp(tM)$ for some (skew-adjoint) essential $M$ such that $-iM$ is not positive.

Suppose furthermore that there is a copy of $\sll(2,\R)$ generated by $\{M,N_+,N_-\}$ inside $\lieg$, i.e. $ [M,N_\pm]=\pm\lambda N_\pm$ and $[N_+,N_-]=\lambda M$ with $\lambda>0$.
If now $N\doteq \frac{1}{2}(N_++N_-)$ generates a compact subgroup of $G$, i.e. $U(\mu(t))=\exp(tN)$ and $\mu(\frac{2\pi}{\lambda})=1$,  then observables in $\A(W(\lambda,x))$ and $\A(\mu(\frac{\pi}{\lambda})W(\lambda,x))$ commute weakly, meaning that 
\begin{equation}
(\Omega,AB\Omega)=(\Omega,BA\Omega)
\end{equation}
for each $A\in \A(W(\lambda,x))$ and $B\in\A(\mu(\frac{\pi}{\lambda})W(\lambda,x))$.
The same holds then also for the wedge pairs $\A(gW(\lambda,x))$ and $\A(g\mu(\frac{\pi}{\lambda})W(\lambda,x))$ for each $g\in G$.
\end{thm}
\begin{proof}
Observe first that $[N,M]=\frac{1}{2}\lambda(N_+-N_-)$ and $[N,\frac{1}{2}\lambda(N_+-N_-)]=-\lambda^2 M$. Hence 
\begin{equation}
\exp(tN)M\exp(-tN)=\exp(\ad(tN))(M)=\cos(\lambda t)M+\frac{1}{2}\sin(\lambda t)(N_+-N_-).
\end{equation}
Setting $t=\frac{\pi}{\lambda}$ and exponentiating, we get
\begin{equation}\label{eq45}
\exp\left(\frac{\pi}{\lambda}N\right)\exp\left(\frac{i\pi}{\lambda}M\right)\exp\left(-\frac{\pi}{\lambda}N\right)=\exp\left(-\frac{i\pi}{\lambda}M\right)
\end{equation}
whereever these operators are defined. Also we know from equation (\ref{eq42}) that
\begin{align*}
J\exp(tN)&=J\slim_{n\to\infty}\left(\exp(tN_+/2)\exp(tN_-/2)\right)^n\\
&=\slim_{n\to\infty}\left(\exp(-tN_+/2)\exp(-tN_-/2)\right)^nJ=\exp(-tN)J.
\end{align*}

Now for all $B\in\A(\mu(\frac{\pi}{\lambda})W(\lambda,x))$ we have \begin{equation}
\exp(\frac{\pi}{\lambda}N)B\exp(-\frac{\pi}{\lambda}N)\in\A(\mu(2\frac{\pi}{\lambda})W(\lambda,x))=\A(W(\lambda,x)).
\end{equation}
Hence we can conclude
\begin{equation}
JB\Omega=J\exp(\frac{\pi}{\lambda}N)\exp(\frac{\pi}{\lambda}N)B\exp(-\frac{\pi}{\lambda}N)\Omega=\exp(\frac{\pi}{\lambda}N)J\exp(\frac{\pi}{\lambda}N)B\exp(-\frac{\pi}{\lambda}N)\Omega
\end{equation}
where we used $\exp(\frac{\pi}{\lambda}N)=\exp(-\frac{\pi}{\lambda}N)$. Going on using equations (\ref{eq41}) and (\ref{eq45}) we get 
\begin{equation}
JB\Omega=\exp(\frac{\pi}{\lambda}N)\exp\left(\frac{i\pi}{\lambda}M\right)\exp(\frac{\pi}{\lambda}N)B^*\Omega=\exp\left(-\frac{i\pi}{\lambda}M\right)B^*\Omega.
\end{equation}
This then finally gives for all $A\in\A(W(\lambda,x))$ and $B\in\A(\mu(\frac{\pi}{\lambda})W(\lambda,x))$ that
\begin{align*}
(\Omega,AB\Omega)&=(\Omega,AJJB\Omega)\\
&=(A^*\Omega, J\exp\left(-\frac{i\pi}{\lambda}M\right)B^*\Omega)\\
&=\overline{(JA^*\Omega, \exp\left(-\frac{i\pi}{\lambda}M\right)B^*\Omega)}\\
&=\overline{(\exp\left(\frac{i\pi}{\lambda}M\right)A\Omega, \exp\left(-\frac{i\pi}{\lambda}M\right)B^*\Omega)}\\
&= (\exp\left(-\frac{i\pi}{\lambda}M\right)B^*\Omega,\exp\left(\frac{i\pi}{\lambda}M\right)A\Omega)\\
&=(\Omega,BA\Omega).
\end{align*}
\end{proof}
\subsection{Concrete examples}
The theorems presented above can be, in particular, applied to the situations of Minkowski space-time, de-Sitter space-time and Anti-de-Sitter space-time, each of them  at least $3$-dimensional, and their corresponding (identity components of the) isometry group. In each of these cases the (skew-adjoint) generator $M$ of a boost subgroup serves as an essential element for which the selfadjoint operator $-iM$ is not positive. The fact that these generators are essential was shown in section \ref{examples}; to see the positivity  observe that because the dimension of the space-times is at least $3$, there is a (skew-adjoint) rotation generator $N$ in the respective Lie algebras that does not commute with $M$. From the concrete commutation relations (\ref{cr}) one has then
\begin{equation}
\exp(\pi N)M\exp(-\pi N)=-M
\end{equation}
and thus $-iM$ can not be positive.
Also one easily sees by direct inspection of the commutation relations (\ref{cr}) that there always is a second boost generator $N'$ such that $\{M,N+N',N-N'\}$ generates a copy of $\sll(2,\R)$ inside $\lieg$. If $M=M_{01}$ then one can for instance pick $N=M_{1j}$ and $N'=M_{0j}$ for $j=2,3,\ldots, n-1$.
In each of these cases the rotation $\frac{1}{2}((N+N')+(N-N'))=N$ generates a compact rotation group.
While the resulting concrete statements for the cases of de-Sitter and Anti-de-Sitter space can be found in \cite{bb99} and \cite{bs04}  as an example we state the then resulting theorem for Minkowski space-time here.
\begin{thm}
Let an algebraic quantum field theory on $n$-dimesnional Minkowski space-time in the form previously presented be given. Let in particular $\Omega$ be a passive, weakly mixing and noncentral state for each observer travelling along the worldline generated by (a conjugate of) a boost subgroup. Then one has
\abc
\begin{enumerate}
\item $\Omega$ is invariant under the whole Poincar\'e group;
\item $\Omega$ is cyclic for each wedge algebra $\A(W)$ with $W=gW_R$ for some $g\in G$;
\item an observer travelling along a worldline generated by a boost subgroup sees $\Omega$ as a $2\pi$-KMS state;
\item the action of modular group for the right wedge $W_R$ coincides with the corresponding boost subgroup action.
\item Observables in a wedge algebra commute with the observables in the opposite wedge algebra weakly; here the opposite wedge of the right wedge is the image of this wedge under a rotation in the $(1,2)$-plane by $\pi$ also called the left wedge $W_L$. In general $gW_R$ has opposite wedge $gW_L$ for $g\in G$.  
\end{enumerate}
\end{thm}
We do not want to forget to mention that such a result (for the case of Minkowski space-time) had also been obtained in \cite{k02} under  different assumptions. There the author shows that if a state  is passive with respect to all generators of  time evolutions of systems that move at arbitrary constant velocities and if in addition the unitaries implementing the translation symmetry belong to the observable algebra, then the spectrum condition holds (\cite[Prop. 5.1.]{k02}) and later he shows (using the spectrum condition, see \cite[Prop. 6.1.]{k02}) that uniformly accelerated observers see the vacuum as a KMS-state at a fixed Unruh-temperature.
\subsection{Further examples using conformally covariant theories}
It is an easy observation that all the previous results remain true if $G$ is assumed to be a subgroup of the conformal group (instead of the isometry group) of the given spacetime manifold $\M$.
In \cite{bms00} the authors show how to construct among other things conformally covariant nets of local algebras on a special class of Robertson-Walker spacetimes. These spacetimes are Lorentzian warped products topologically equivalent to $\R\times S^3$ where the metric in the usual cylindrical coordinates $(t,\chi,\theta,\phi)$ is of the form
$$
ds^2=dt^2-S(t)^2\left(d\chi^2+\sin^2(\chi)\left(d\theta^2+\sin^2(\theta)d\phi^2\right)\right).
$$
Here $S$ is assumed to be a positive, smooth (warping) function. Then following \cite{he73} one can define a new time variable $\tau$ via
$$
\frac{d\tau}{dt}=\frac{1}{S(t)}
$$
and in these new coordinates the metric takes the form
$$
ds^2=S^2(\tau)\left(d\tau^2-d\chi^2+\sin^2(\chi)\left(d\theta^2+\sin^2(\theta)d\phi^2\right)\right).
$$
Now $\tau$ has, as a strictly increasing continuous function of $t$, some open interval as its range.
In the special case when this range is of the form $\left(-\frac{\pi}{2},\frac{\pi}{2}\right)$ the corresponding Robertson-Walker spacetime has $\SO(4,1)$ as its conformal group and is conformally isomorphic to four-dimensional de-Sitter spacetime. In \cite{bms00} a method called {\it transplantation} is then used to construct conformally covariant nets on such a Robertson-Walker spacetime.
As discussed in section \ref{examples} the group $\SO(4,1)$ has essential elements and due to the conformal equivalence to the de-Sitter case all our results are applicable for these special Robertson-Walker spaces as well.

\section{Conclusion}
The results given propose  a unified treatment for the cases considered in \cite{bb99} and \cite{bs04}. In addition, it also covers the case of Minkowski space-time in at least $3$ dimensions (under assumptions different from those  in \cite{k02}). In general the results show that the requirement of a state to be stable (in the sense of passivity) for a certain class of observers is very restrictive and selects a very special class of states having desirable properties such as invariance and the Reeh-Schlieder property.

It would be desirable to find further examples of space-times fitting into the presented framework. The author's attempts to incorporate also Einstein's static universe or other Robertson-Walker space-times failed so far, due to the lack of essential generators in the corresponding isometry groups. To this end a deeper analysis of the condition of essentiality seems necessary. In particular, the task of classifying Lie groups having essential generators could be interesting to investigate. 
\section{Acknowledgements}
The author would like to thank S. J. Summers and D. Buchholz for initiating the work that led to this paper and for helpful discussions throughout the process of writing it. 

\end{document}